\documentclass{amsart}
\begin{document}
\baselineskip=15pt
\newtheorem{theorem}{Theorem}
\newtheorem{lemma}{Lemma}

\title[Exact solutions to the compressible Navier-Stokes]
{Exact solutions to the compressible Navier-Stokes equations with
the Coriolis and friction terms}


\author{Anastasya Korshunova}

\address{Department of Differential Equations \& Mechanics and Mathematics Faculty,
Moscow State University, Moscow, 119992,
 Russia}

\email{korshunova\underline{\phantom{a}}aa@mail.ru}




\begin{abstract}
We consider special solution to the 3D Navier-Stokes system with and
without the Coriolis force and dry friction and find the respective
initial data implying a finite time gradient catastrophe. The paper
can be considered as extension of the results \cite{Shiv}.
\end{abstract}

\maketitle

We consider the following gas-dynamic like system:

\begin{equation}
\label{gas_syst1}\dfrac{\partial \rho}{\partial
t}+(\textbf{v},\nabla)\rho+\rho(\nabla,\textbf{v})=0,
\end{equation}

\begin{equation}
\label{gas_syst2} \rho(\dfrac{\partial\textbf{v}}{\partial
t}+(\textbf{v},\nabla)\textbf{v})=-\nabla p+\rho
(L\mathbf{v}+\mu\Delta\mathbf{v}),
\end{equation}

\begin{equation}
\label{gas_syst3}\dfrac{\partial p}{\partial t}+(\textbf{v},\nabla
p)+\gamma p \,{\rm div} \textbf{v}=0,
\end{equation}

\hspace*{-0,7cm} where $\rho(t,\textbf{x})$, $p(t,\textbf{x})$,
$\textbf{v}(t,\textbf{x})= (v_1, v_2, v_3)$ are density, pressure
and velocity vector, respectively, $\mu$ is the dynamic viscosity
coefficient and
\begin{center}
$$ L=\begin{pmatrix} -m & -l & 0\\
l & -m & 0\\ 0 & 0 & 0
\end{pmatrix}, $$
\end{center}
\hspace*{-0,6cm}

$m=const \geq 0$ is the friction coefficient, $l=const$ is the
Coriolis parameter, $\textbf{x}\in {\mathbb R}^n, \, t\ge 0.$

Exact solutions of the system (\ref{gas_syst1})-(\ref{gas_syst3})
has been an area of intensive research activity in the last decades
(e.g. \cite{Sidorov1}, \cite{Sidorov2}, \cite{Sidorov3},
\cite{Sidorov4}, \cite{Sedov}). Examples of exact solutions with a
special initial distribution of the tangential component are given
in ~\cite{Ovsyannikov}.

Below we consider the solution to
(\ref{gas_syst1})-(\ref{gas_syst3}) in several particular cases.

\section{Case $l=0$, $m=0$ (without the Coriolis force and dry friction)}
\label{d1}

Let us consider the velocity field and the density in the following
form:

\begin{equation}
\label{density}\textbf{v}=\begin{pmatrix} \alpha(t)x\\
\beta (t)y\\W(x,t)
\end{pmatrix},\quad\quad \rho(t,x)=\sigma(t)+\dfrac{\rho_0}{U}(\alpha(t)+\beta (t))x.
\end{equation}

Here $\alpha (t)+\beta (t)$ is the divergency of velocity field.

Firstly we are going to define the functions $\rho(x,t)$, $\alpha
(t)$ and $\beta (t)$.

For this class of solution the conversation of mass
(\ref{gas_syst1}) and vorticity conversation equation

\begin{equation}
\label{vorticity}\nabla\times[\rho(\dfrac{\partial\textbf{v}}{\partial
t}+(\textbf{v},\nabla)\textbf{v})-\mu\triangle\textbf{v}]=0,
\end{equation}

\hspace*{-0,6cm}which follows from (\ref{gas_syst2}), gives

\begin{equation}
\label{mass}\dot{\sigma}+\dfrac{\rho_0}{U}(\dot{\alpha
}+\dot{\beta })x+\alpha \dfrac{\rho_0}{U}x(\alpha +\beta
)+[\sigma+\dfrac{\rho_0}{U}(\alpha +\beta )x](\alpha +\beta ),
\end{equation}

$$[\sigma+\dfrac{\rho_0}{U}(\alpha +\beta
)x]\left[\dfrac{\partial^2 W}{\partial x\partial t}+\alpha
x\dfrac{\partial^2 W}{\partial
x^2}\right]+\dfrac{\rho_0}{U}(\alpha +\beta )\left[\dfrac{\partial
W}{\partial t}+\alpha x\dfrac{\partial W}{\partial x}\right]=$$

\begin{equation}
\label{vorticity}=-\alpha \dfrac{\partial W}{\partial
x}\left[\sigma+\dfrac{\rho_0}{U}(\alpha +\beta
)x\right]+\mu\dfrac{\partial^3 W}{\partial x^3},
\end{equation}

$$(\alpha +\beta )(\dot{\beta }+\beta ^2)=0.$$

In~\cite{Shiv} it was shown that $\alpha (t)$, $\beta (t)$ and
$\sigma(t)$ satisfy the following system:

\begin{equation}
\label{A}\beta(t)=\dfrac1{t+B},\quad -\alpha(t)=c(t)+\beta(t),
\end{equation}

\begin{equation}
\label{C_t}c(t)=\dfrac{{exp}\left(-\int\limits_0^t
\alpha(\tau)d\tau\right)}{C+\int\limits_0^t{exp}\left(\int\limits_0^t
\alpha(\tau')d\tau'\right)d\tau},
\end{equation}

\hspace*{-0,6cm}where $\sigma(t)={exp}\left(\int\limits_0^t
c(\tau)d\tau\right)$ and the constants $B$ and $C$ can be found
from the initial data, namely: $B=(\beta (0))^{-1}$,
$C=(c(0))^{-1}=-(\alpha (0)+\beta (0))^{-1}$.

From (\ref{A}) and (\ref{C_t}) we find functions $\alpha(t)$,
$\beta(t)$ and $c(t)$. Substituting $-\alpha(t)$ for
$c(t)+\dfrac1{t+B}$ in (\ref{C}), we obtain:

\begin{equation}
\label{int_c} c(t)=\dfrac{(t+B){exp}\left(\int\limits_0^t
c(\tau)d\tau\right)}{C-\int\limits_0^t(\tau+B){exp}\left(\int\limits_0^{\tau}
c(\tau')d\tau'\right)d\tau},
\end{equation}

Let us denote
$p(t)=C-\int\limits_0^t(\tau+B){exp}\left(\int\limits_0^{\tau}
c(\tau')d\tau'\right)d\tau$. It is easy to see that

$$c(t)=\left(\ln\Big|\dfrac{p'(t)}{t+B}\Big|\right)'.$$

Thus, (\ref{int_c}) implies:

\begin{equation}
\label{equ_p} \dfrac{p'(t)}{t+B}=\dfrac{C_1}{p(t)},
\end{equation}

therefore $p^2(t)=C_1(t^2+2Bt+C_2)$. Taking into account (3), we
obtain:

$$c(t)=-\dfrac{p'(t)}{p(t)}=-\dfrac{C_1(t+B)}{p^2(t)}=-\dfrac{t+B}{t^2+2Bt+C_2},$$

\hspace*{-0,6cm}where $C_1=-\dfrac{C}{B}$, $C_2=-BC$.

Therefore, we can find all functions:

\begin{equation}
\label{equ_rho1}\alpha(t)=-\dfrac1{t+B}+\dfrac{t+B}{t^2+2Bt+C_2},
\quad \beta(t)=\dfrac1{t+B},
\end{equation}

\begin{equation}
\label{equ_sigma}\sigma(t)=\dfrac{C_2^{1/2}}{\sqrt{t^2+2Bt+C_2}},
\end{equation}

\begin{equation}
\label{equ_rho2}
\rho(x,t)=\sigma(t)-\dfrac{\rho_0}{U}c(t)x=\dfrac{C_2^{1/2}}{\sqrt{t^2+2Bt+C_2}}+\dfrac{\rho_0}{U}\dfrac{t+B}{t^2+2Bt+C_2}x
\end{equation}

Having the analytic form of the solution, we can find initial data
implying an unbounded increasing of its derivative in a finite time
(gradient catastrophe). It is clear that it is sufficient to find
when the $f(t)=$ $=(t+B)(t^2+2Bt-BC)$ has the positive zeroes. We
have the following results:

1) if $B > 0$, $C\leq 0$ ($\beta (0) > 0$, $\alpha (0) \geq -\beta
(0)$) then $f(t)\neq 0$, $\forall t>0$;

 2) $B > 0$, $C > 0$ ($\beta (0) > 0$, $\alpha (0)
> -\beta (0)$) then $f(t)$ has one positive zero:

$$t=-B+\sqrt{B^2+BC}=\dfrac1{\beta (0)}\left(\sqrt{\dfrac{\alpha
(0)}{\alpha (0)+\beta (0)}}-1\right);$$

3) if $B < 0$, $C < 0$ or $C > -B$ ($\beta (0) < 0$, $\alpha (0)
> 0$, $\alpha (0)\neq \beta (0)$)then $f(t)$ has one positive
zero $t=-B$;

4) if $C = 0$, $B < 0$ ($\beta (0) < 0$, $\alpha (0) = -\beta
(0)$), then $f(t)=(t+B)(t^2+2Bt)$. Thus, $f(t)=0$ in the points
$t=T_1=-B$ and $t=T_2=-2B$;

5) if $B < 0$, $0 < C < -B$ ($\beta (0) < 0$, $\alpha (0) < 0$),
then $f(t)$ turns into zero in three points:

$$t=T_1=-B-\sqrt{B^2+BC}=\dfrac1{\beta (0)}\left(\sqrt{\dfrac{\alpha
(0)}{\alpha (0)+\beta (0)}}-1\right),$$

$$t=T_2=-B,$$

$$t=T_3=-B+\sqrt{B^2+BC}=-\dfrac1{\beta (0)}\left(\sqrt{\dfrac{\alpha
(0)}{\alpha (0)+\beta (0)}}+1\right).$$

It is obviously that $T_1 < T_2 < T_3$.

Now we are ready to find the third component of velocity $W(t,x)$.

According to~\cite{Shiv} $W(t,x)$ solves the PDE:

\begin{equation}
\label{equ_W} \rho\left(\dfrac{\partial W}{\partial t}+\alpha
(t)x\dfrac{\partial W}{\partial x}\right)=\mu\dfrac{\partial^2
W}{\partial x^2}
\end{equation}

We consider $W(x,t)=W(\omega(t)x-\lambda(t))=:W(f)$. Thus, we
have:

$$\dfrac{\partial W}{\partial t}=(\dot{\omega}(t)x-\dot{\lambda}(t))W'(f),$$

$$\dfrac{\partial W}{\partial x}=\omega(t)W'(f),\quad \dfrac{\partial^2 W}{\partial
x^2}=\omega^2(t)W''(f).$$

Using the specific form of $\rho(x,t)$ (\ref{density}), we obtain
from (\ref{equ_W}):
\begin{equation}
\label{equ_W2}(\sigma(t)+g(t)x)((\dot{\omega}(t)-a(t)\omega(t))x-\dot{\lambda}(t))W'(f)=\mu\omega^2(t)W''(f).
\end{equation}

Equation (\ref{equ_W2}) implies:
\begin{equation}
\label{equ_l_w}
\omega^{-2}(t)(\sigma(t)+g(t)x)((\dot{\omega}(t)-a(t)\omega(t))x-\dot{\lambda}(t))=(\omega(t)x-\lambda(t))^2
\end{equation}

Under this condition we find $\omega(t)$ and $\lambda(t)$. From
(\ref{equ_l_w}) we obtain:
\begin{equation}
\label{system_l_w1}g(t)(\dot{\omega}(t)-a(t)\omega(t))=\omega^4(t),
\end{equation}
\begin{equation}
\label{system_l_w2}
-\dot{\lambda}(t)g(t)+\sigma(t)(\dot{\omega}(t)-a(t)\omega(t))=-2\omega^3(t)\lambda(t),
\end{equation}
\begin{equation}
\label{system_l_w3}-\dot{\lambda}(t)\sigma(t)=\lambda^2(t)\omega^2(t).
\end{equation}

In this system $\omega(t)$ and $\lambda(t)$ are unknown and
$\sigma(t)=\dfrac{C_2^{1/2}}{\sqrt{t^2+2Bt+C_2}}$,
$g(t)=\dfrac{\rho_0}{U}\dfrac{t+B}{t^2+2Bt+C_2}$ (See
(\ref{equ_sigma}) and (\ref{equ_rho2})).

Using (\ref{system_l_w2}) and (\ref{system_l_w3}), we find
\begin{equation}
\label{cond_l_w}
\lambda(t)=-\dfrac{\sigma(t)}{g(t)}\omega(t)=-\dfrac{UC_2^{1/2}}{\rho_0}\dfrac{\sqrt{t^2+2Bt+C_2}}{t+B}\omega(t).
\end{equation}

Let us denote $s(t)=\omega^{-3}(t)$. Thus, from (\ref{system_l_w1})
we get

$$\dot{s}(t)+3a(t)s(t)=-3g^{-1}(t).$$

It can be readily concluded that

\begin{equation}
\label{sol_w}
\omega(t)=\dfrac{C_2^{1/2}(t+B)}{B\sqrt{t^2+2Bt+C_2}}\left(F(t)+C_3\right)^{-1/3},
\end{equation}

\hspace*{-0,6cm}where
\begin{equation}
\label{F}F(t)
=-\dfrac12K_1(t+B)\sqrt{t^2+2Bt+C_2}+\dfrac12K_1(C_2-B^2)\ln|t+B+\sqrt{t^2+2Bt+C_2}|
\end{equation}

\hspace*{-0,6cm}and

$$K_1=-3\dfrac{UC_2^{3/2}}{B^3\rho_0},\quad C_3=\omega(0)-F(0).$$

From (\ref{cond_l_w}) we find
\begin{equation}
\label{sol_l}\lambda(t)=K_2\left(F(t)+C_3\right)^{-1/3},
\end{equation}

\hspace*{-0,6cm}where $K_2=-\dfrac{UC_2^{1/2}K_1}{\rho_0}$.

Finally we show that (\ref{system_l_w1})-(\ref{system_l_w3}) are
compatible. Substituting $\omega(t)$ and $\lambda(t)$ (see
(\ref{sol_w}) and (\ref{sol_l})) in (\ref{system_l_w3}), we
obtain
\vspace*{-0,6cm}

\begin{equation}
\label{prov}\dfrac{K_2}{3}\left(F(t)+C_3\right)^{-4/3}F'(t)\dfrac{C_2^{1/2}}{q(t)}=K_2^2\left(F(t)+C_3\right)^{-4/3}\dfrac{C_2(t+B)^2}{B^2q^2(t)}
\end{equation}

\hspace*{-0,6cm}(here $q(t)=$ $=\sqrt{t^2+2Bt+C_2}$).

Further, (\ref{F}) gets

$$F'(t)=-\dfrac12 K_1 q(t)-\dfrac12
K_1\dfrac{(t+B)^2}{q(t)}+\dfrac12
K_1(C_2-B^2)\dfrac1{q(t)}=-\dfrac{K_1(t+B)^2}{q(t)}.$$

Therefore, it is easy to see that (\ref{prov}) holds identically.

For $\omega(t)$ and $\lambda(t)$ defined from (\ref{equ_l_w}) we
have $\mu W''(f)=f^2W'(t)$. Integrating gives

$$W(f)=W'(0)\int\exp\left(\dfrac1{3\mu}f^3\right)\,df,$$

\hspace*{-0,6cm}where $f(x,t)=\omega(t)x-\lambda(t)$.

\section{Case $l\neq 0$, $m\neq 0$}
\label{d2}

We will find the velocity vector and the density in the following
form:
\begin{equation}
\label{velocity}\textbf{v}=\begin{pmatrix} \alpha (t)x+\gamma(t)y\\
\xi(t)x+\beta (t)y\\W(x,y,t)
\end{pmatrix}, \quad \rho(t,x)=\sigma(t)+g(t)x=\sigma(t)+\dfrac{\rho_0}{U}(\alpha
(t)+\beta (t))x.
\end{equation}

Here the divergency is $\alpha(t)+\beta(t)$, the vorticity is
$\left(\dfrac{\partial W}{\partial y}, -\dfrac{\partial
W}{\partial x}, \xi (t)+\gamma (t)\right)^T$.

Firstly we find the functions $\rho(x,t)$, $\alpha(t)$,
$\beta(t)$, $\gamma (t)$ and $\xi (t)$.

The conservation of mass (\ref{gas_syst1}) yields:

$$\dot{\sigma}+\dot{g}x+(\alpha x+\gamma y)g+(\sigma+gx)(\alpha +\beta )=0.$$

Therefore

\begin{equation}
\label{mass-conv1}\dot{g}+\alpha g+g(\alpha +\beta )=0,
\end{equation}

\begin{equation}
\label{mass-conv2} \dot{\sigma}+\sigma(\alpha +\beta )=0,
\end{equation}

\begin{equation}
\label{mass-conv3}g\gamma =0.
\end{equation}

From the vorticity conservation equation (\ref{vorticity}) we have
the following equations:

$$\rho\left(\frac{\partial^2{W}}{\partial t\partial y}+\gamma
\frac{\partial{W}}{\partial x}+(\alpha x+\gamma
y)\frac{\partial^2{W}}{\partial x\partial y}+\beta
\frac{\partial{W}}{\partial y}+(\xi x+\beta
y)\frac{\partial^2{W}}{\partial y^2}\right)=$$

\begin{equation}
\label{W(x,y)1}=\mu\left(\frac{\partial^3{W}}{\partial x^2\partial
y}+\frac{\partial^3{W}}{\partial y^3}\right),
\end{equation}

$$\rho\left(\frac{\partial^2{W}}{\partial t\partial x}+\alpha
\frac{\partial{W}}{\partial x}+(\alpha x+\gamma
y)\frac{\partial^2{W}}{\partial x^2}+\xi
\frac{\partial{W}}{\partial y}+(\xi x+\beta
y)\frac{\partial^2{W}}{\partial x\partial y}\right)+$$

\begin{equation}
\label{W(x,y)2}+\frac{\partial{\rho}}{\partial
x}\left(\frac{\partial{W}}{\partial t}+(\alpha x+\gamma
y)\frac{\partial{W}}{\partial x}+(\xi x+\beta
y)\frac{\partial{W}}{\partial y}\right)=
-\mu\left(\frac{\partial^3{W}}{\partial
x^3}+\frac{\partial^3{W}}{\partial x\partial y^2}\right),
\end{equation}

$$\rho\left(\dot{\xi }-\dot{\gamma}+(\alpha +\beta )(\xi -\gamma )-l(\alpha +\beta )+m(\xi -\gamma )\right)+$$

\begin{equation}
\label{W(x,y)3}+\frac{\partial{\rho}}{\partial x}\left(\dot{\xi
}x+\dot{\beta }y+(\beta +m)(\xi x+\beta y)+(\xi -l)(\alpha
x+\gamma y)\right)=0.
\end{equation}

From (\ref{W(x,y)3}) we get
\begin{equation}
\label{syst_v1}g(2\dot{\xi }-\dot{\gamma}+(\alpha +\beta +m)(2\xi
-\gamma )-2l\alpha -l\beta )=0,
\end{equation}

\begin{equation}
\label{syst_v2}g(\dot{\beta }+\beta ^2+m\beta +(\xi -l)\gamma )=0,
\end{equation}

\begin{equation}
\label{syst_v3}\sigma(\dot{\xi}-\dot{\gamma}+(\alpha +\beta
+m)(\xi -\gamma )-l\alpha -l\beta )=0.
\end{equation}

Below we treat particular cases separately.

1. $g(t)\not\equiv 0$, $\sigma(t)\not\equiv 0$. It follows from
(\ref{mass-conv3}) that $\gamma (t)\equiv 0$. In this case instead
of system (\ref{syst_v1})-(\ref{syst_v3}) we have:

\begin{equation}
\label{syst_vel1}2\dot{\xi}+2(\alpha+\beta)\xi -2l\alpha+2m\xi
-l\beta =0,
\end{equation}

\begin{equation}
\label{syst_vel2}\dot{\beta }+\beta ^2+m\beta =0,
\end{equation}

\begin{equation}
\label{syst_vel3}\dot{\xi}+(\alpha +\beta )\xi -l\alpha +m\xi
-l\beta =0.
\end{equation}

From (\ref{syst_vel1}) and (\ref{syst_vel3}) we get immediately that
$\beta (t)\equiv 0$. To find functions $\alpha (t)$, $\xi (t)$ and
$\sigma(t)$ we have the system of differential equations obtained
from (\ref{mass-conv1}), (\ref{mass-conv2}) and (\ref{syst_vel1}):

$$\dot{\alpha }+2\alpha ^2=0,$$

$$\dot{\sigma}+\alpha \sigma=0, $$

$$\dot{\xi}+(\alpha +m)\xi -l\alpha =0.$$

Integrating gives

\begin{equation}
\label{alpha}\alpha (t)=\dfrac1{2t+K_1},
\end{equation}

\hspace*{-0,6cm}where $K_1=(\alpha (0))^{-1}$,

\begin{equation}
\label{alpha2}\sigma(t)=\dfrac{K_2}{\sqrt{|2t+K_1|}},
\end{equation}

\hspace*{-0,6cm}where
$K_2=\sigma(0)\sqrt{|K_1|}=\sigma(0)\sqrt{|(\alpha (0))^{-1}|}$,

\begin{equation}
\label{xi}\xi (t)=C(t)\dfrac{e^{-mt}}{\sqrt{|2t+K_1|}},
\end{equation}

\hspace*{-0,6cm}where
$C(t)=K_3+l\int\limits_0^t\dfrac{e^{m\tau}}{\sqrt{|2\tau+K_1|}}\,d\tau$.

If $m=0$, then we can integrate (\ref{xi}):

\begin{equation}
\label{xi0}\xi (t)=\dfrac{K_4}{\sqrt{|2t+K_1|}}+l,
\end{equation}

\hspace*{-0,6cm}where $K_4=(\xi (0)-l)\sqrt{|K_1|}=(\xi
(0)-l)\sqrt{|(\alpha (0))^{-1}|}$.

Thus, for $g(t)\not\equiv 0$, $\sigma(t)\not\equiv 0$ and $m=0$ we
find the following solution for $\rho(x,t)$ and
$\mathbf{v}(t,x,y)$:

\begin{equation}
\label{r0}\rho(t,x)=\dfrac{\sigma(0)}{\alpha (0)\sqrt{|2t+(\alpha
(0))^{-1}|}}+\dfrac{\rho_0}{U}\dfrac{x}{2t+(\alpha (0))^{-1}},
\end{equation}

\begin{equation}
\label{v0_1}\mathbf{v}_1=\alpha
(t)x+\gamma(t)y=\dfrac{x}{2t+(\alpha (0))^{-1}},
\end{equation}

\begin{equation}
\label{v0_2}\mathbf{v}_2=\xi(t)x+\beta(t)y=\left(l+\dfrac{(\xi
(0)-l)(\alpha (0))^{-1/2}}{\sqrt{|2t+(\alpha (0))^{-1}|}}\right)x.
\end{equation}

2. $g(t)\not\equiv 0$, $\sigma(t)\equiv 0$. From
(\ref{mass-conv3}) we get $\gamma (t)\equiv 0$. From
(\ref{syst_v2}) we find $\beta (t)$ as follows:

\begin{equation}
\label{beta}\beta (t)=\begin{cases} 0, &\text{если $\beta (0)=0$;}\\
\dfrac{mC_1}{e^{mt}-C_1},&\text{если $\beta (0)\neq 0$ и $m\neq 0$;}\\
\dfrac1{t+(\beta (0))^{-1}}, &\text{если $\beta (0)\neq 0$ и
$m=0$;}\end{cases}
\end{equation}

\hspace*{-0,6cm}where $C_1=\dfrac{b(0)}{m+b(0)}$.

It is easy to see from (\ref{mass-conv1})-(\ref{mass-conv3}) and
(\ref{syst_v1})-(\ref{syst_v3}) that functions $\alpha (t)$ and $\xi
(t)$ solve the following system:

\begin{equation}
\label{a_and_xi1}\dot{\xi}+(\alpha +\beta )\xi -l\alpha +m\xi
-\dfrac l2\beta =0,
\end{equation}

\begin{equation}
\label{a_and_xi2}\dot{\alpha }+\dot{\beta }+\alpha (\alpha +\beta
)+(\alpha +\beta )^2=0.
\end{equation}

Further, assume $m=0$. Using (\ref{syst_v2}) and
(\ref{a_and_xi1})-(\ref{a_and_xi2}), we obtain

\begin{equation}
\label{a_&_x1}\dot{\xi}+(\alpha +\beta )\xi -l\alpha -\dfrac
l2\beta =0,
\end{equation}

\begin{equation}
\label{a_&_x2}\dot{\alpha }+2\alpha ^2+3\alpha \beta =0.
\end{equation}

If $\alpha (0)=0$ then the solution of (\ref{a_&_x2}) is zero
identically. The function $\xi (t)$ can be found from
(\ref{a_&_x1}). Therefore

\begin{equation}
\label{xi2}\xi (t)=\dfrac{lt+C_3}{2(t+C_2)},
\end{equation}

\hspace*{-0,6cm}where

\begin{equation}
\label{C}C_2=(\beta (0))^{-1},\quad C_3=2(\beta (0))^{-1}\xi (0).
\end{equation}

Thus, in this case we obtain

\begin{equation}
\label{r}\rho(t,x)=\dfrac{\rho_0x}{U(t+C_2)},
\end{equation}

\begin{equation}
\label{v}\mathbf{v}_1=0,\quad
\mathbf{v}_2=\dfrac{lt+C_3}{2(t+C_2)}x+\dfrac1{t+C_2}y,
\end{equation}

\hspace*{-0,6cm}where the constants $C_2$ и $C_3$ are determined
early (see (\ref{C})).

If $\alpha (0)\neq 0$, we denote $A(t)=(\alpha (t))^{-1}$ and we
get from (\ref{a_&_x2}):

$$\dot{A}(t)-3\beta (t)A(t)-2=0.$$

Therefore, we can find

\begin{equation}
\label{alpha2}\alpha (t)=\dfrac1{(t+C_2)(C_4(t+C_2)^2-1)},
\end{equation}

\hspace*{-0,6cm}where

\begin{equation}
\label{C2}C_2=(\beta (0))^{-1},\quad C_4=\dfrac{(\alpha
(0))^{-1}+C_2}{C_2^3}.
\end{equation}

Then we can find from (\ref{a_&_x1})

\begin{equation}
\label{xi3}\xi(t)=\dfrac{C_5}{\sqrt{|C_4(t+C_2)^2-1|}}+\dfrac l2,
\end{equation}

\hspace*{-0,6cm}where

\begin{equation}
\label{C3}C_5=(\xi (0)-\dfrac l2)\sqrt{|C_4C_2^2-1|}.
\end{equation}

Thus, if $g(t)\not\equiv 0$, $\sigma(t)\equiv 0$, $m=0$ and $\alpha
(0)\neq0$ we have

\begin{equation}
\label{r2}\rho(t,x)=\dfrac{\rho_0}{U}\dfrac{C_4(t+C_2)^2}{(t+C_2)(C_4(t+C_2)^2-1)}x,
\end{equation}

\begin{equation}
\label{v2_1}\mathbf{v}_1=\dfrac{x}{(t+C_2)(C_4(t+C_2)^2-1)},
\end{equation}

\begin{equation}
\label{v2_2}\mathbf{v}_2=\left(\dfrac{C_5}{\sqrt{|C_4(t+C_2)^2-1|}}+\dfrac
l2\right)x+\dfrac1{t+C_2}y.
\end{equation}

\hspace*{-0,6cm}where constants $C_2$, $C_4$ и $C_5$ are
determined in (\ref{C2}) and (\ref{C3}).

\subsection{The case $W(x,y,t)=\omega_1(t)x+\omega_2(t)y+\omega_3(t)$}

Early we find the first and the second component of velocity
vector. In this section we consider the special form of the third
component:

$$W(x,y,t)=\omega_1(t)x+\omega_2(t)y+\omega_3(t).$$

Here (\ref{W(x,y)1}) and (\ref{W(x,y)2}) imply:

\begin{equation}
\label{lin_W(x,y)1}(\sigma+gx)(\dot{\omega}_2+\gamma
\omega_1+\beta \omega_2)=0,
\end{equation}

\begin{equation}
\label{lin_W(x,y)2}g\left(\dot{\omega}_1x+\dot{\omega}_2y+\dot{\omega}_3+(\alpha
x+\gamma y)\omega_1+(\xi x+\beta
y)\omega_2\right)+(\sigma+gx)(\dot{\omega}_1+\alpha \omega_1+\xi
\omega_2)=0
\end{equation}

We treat particular cases separately:

1. $g(t)\not\equiv 0$, $\sigma(t)\not\equiv 0$. From
(\ref{lin_W(x,y)1}) and (\ref{lin_W(x,y)2}) we obtain:

\begin{equation}
\label{lin_W(x,y)1_2}\dot{\omega}_1+\alpha \omega_1+\xi
\omega_2=0,
\end{equation}

\begin{equation}
\label{lin_W(x,y)2_2}\dot{\omega}_2+\beta \omega_2+\gamma
\omega_1=0,
\end{equation}

\begin{equation}
\label{lin_W(x,y)3_2}\dot{\omega}_3=0,
\end{equation}

Early we proved that in the case $g(t)\not\equiv 0$,
$\sigma(t)\not\equiv 0$ we have $\beta (t)\equiv 0$ and $\gamma
(t)\equiv 0$ (see (\ref{v0_1}) and (\ref{v0_2})). Thus, from
(\ref{lin_W(x,y)2_2}) and (\ref{lin_W(x,y)3_2}) one gets

$$\omega_i(t)\equiv\omega_i(0),\quad i=2,3.$$

Further, we can find from (\ref{lin_W(x,y)1_2})

$$\omega_1(t)=K(t)\dfrac{1}{\sqrt{|2t+K_1|}},$$

$$K(t)=\dfrac{l\omega_2(0)}{m}e^{-mt}\int\dfrac{e^{mt}}{\sqrt{|2t+K_1|}}\,dt-l\sqrt{|2t+K_1|}+K_5.$$

\hspace*{-0,6cm}where $K_1=(\alpha (0))^{-1}$.

If $m=0,$ then

$$K(t)=-\omega_2(0)K_4t-\dfrac13l\omega_2(0)|2t+K_1|^{3/2}+K_6,$$

\hspace*{-0,6cm}where $K_4=(\xi (0)-l)\sqrt{|K_1|}$,
$K_6=-(\omega_1(0)+\dfrac13l\omega_2(0)K_1)\sqrt{|K_1|}$.

2. $g(t)\not\equiv 0$, $\sigma(t)\equiv 0$. Then functions
$\omega_i(t)$, $i=1,2,3$ solves the system
(\ref{lin_W(x,y)1_2})-(\ref{lin_W(x,y)3_2}). It is obvious that
$\omega_3(t)\equiv\omega_3(0)$. Functions $\omega_1(t)$ and
$\omega_2(t)$ we find for case $m=0$. We have from
(\ref{mass-conv3}) and (\ref{beta}) $\gamma (t)\equiv 0$, $\beta
(t)=\dfrac1{t+(\beta (0))^{-1}}$. Thus, it is easy to see that

$$\omega_2(t)=\dfrac{\omega_2(0)}{t+C_2},$$

\hspace*{-0,6cm}where $C_2=(\beta (0))^{-1}$.

Further, we can find $\omega_1(t).$ Namely:

a). If $\alpha (0)=0$, then $\alpha (t)\equiv 0$ and $\xi
(t)=\dfrac{lt+C_3}{2(t+C_2)}$, where $C_3=2C_2\xi (0)$,
$C_2=(\beta (0))^{-1}$ (see (\ref{xi2})and (\ref{C})). Therefore,
we obtain from (\ref{lin_W(x,y)1_2}):

\begin{equation}
\label{omega1}\omega_1(t)=-(\dfrac12l\omega_2(0)t+C_6\ln|t+C_2|+C_7),
\end{equation}

\hspace*{-0,6cm}where $C_6=\omega_2(0)(\xi (0)-\dfrac12)\beta
^{-1}(0)$, $C_7=-C_6\ln|C_2|-\omega_1(0)$.

b). If $\alpha (0)\neq0$, then it follows from (\ref{alpha2}) that

\begin{equation}
\label{int}-\int\limits_0^t \alpha
(\tau)\,d\tau=\ln\dfrac{|t+C_2|}{\sqrt{|C_4(t+C_2)^2-1|}}.
\end{equation}

\hspace*{-0,6cm}where $C_2=(\beta (0))^{-1}$, $C_4=\dfrac{(\alpha
(0))^{-1}+C_2}{C_2^3}$.

Using (\ref{xi3}), (\ref{C3}) and (\ref{int}), we obtain:

$$\omega_1(t)=K(t)\dfrac{t+C_2}{\sqrt{|C_4(t+C_2)^2-1|}},$$

\hspace*{-0,6cm}where

$$K(t)=C_8-\dfrac12C_4\arcsin\dfrac1{\sqrt{|C_4|}(t+C_2)}-\dfrac{\sqrt{|C_4(t+C_2)^2-1|}}{2(t+C_2)^2},$$

$$C_8=\sqrt{|C_4C_2^2-1|}\beta (0)(\omega_1(0)+\beta(0))+\dfrac12C_4\arcsin\dfrac{\beta
(0)}{\sqrt{|C_4|}}.$$

We find all components of the velocity field (\ref{velocity}),
where $W(x,y,t)=\omega_1(t)x+$ $+\omega_2(t)y+\omega_3(t)$.
Therefore, we are ready to study the problem of the gradient
catastrophe:

1. If $\alpha (0)\neq 0$, $\beta (0)= 0$ and $\gamma (0)= 0$ then

$$\rho(t,x)=\dfrac{\sigma(0)}{\alpha (0)\sqrt{|2t+K_1|}}+\dfrac{\rho_0}{U}\dfrac{x}{2t+K_1},$$

$$\mathbf{v}_1=\dfrac{x}{2t+K_1},\qquad
\mathbf{v}_2=\left(l+\dfrac{K_4}{\sqrt{|2t+K_1|}}\right)x,$$

$$\mathbf{v}_3=\left(\dfrac{K_6-\omega_2(0)K_4t}{\sqrt{|2t+K_1|}}-\dfrac13l\omega_2(0)(2t+K_1)\right)x+\omega_2(0)y+\omega_3(0);$$

It is obvious that for such velocity field the gradient catastrophe
takes place in the time $t=-\dfrac1{2\alpha (0)}$ if $\alpha (0)<0$.

2. If $\sigma(0)=0$, $\alpha (0)= 0$, $\beta (0)\neq 0$ and
$\gamma (0)= 0$ then

$$\rho(t,x)=\dfrac{\rho_0x}{U(t+C_2)},$$

$$\mathbf{v}_1\equiv 0,\qquad\mathbf{v}_2=\dfrac{lt+C_3}{2(t+C_2)}x+\dfrac1{t+C_2}y,$$

$$\mathbf{v}_3=-(\dfrac12l\omega_2(0)t+C_6\ln|t+C_2|+C_7)x+\dfrac{\omega_2(0)}{t+C_2}y+\omega_3(0);$$

In this case the necessary condition of gradient catastrophe is
$\beta (0)<0$ and its time is equal $t=-(\beta (0))^{-1}$.

3. If $\sigma(0)=0$, $\alpha (0)\neq 0$, $\beta (0)\neq 0$ and
$\gamma (0)= 0$ then

$$\rho(t,x)=\dfrac{\rho_0}{U}\dfrac{C_4(t+C_2)^2}{(t+C_2)(C_4(t+C_2)^2-1)}x,$$

$$\mathbf{v}_1=\dfrac{x}{(t+C_2)(C_4(t+C_2)^2-1)},$$

$$\mathbf{v}_2=(\dfrac{C_5}{\sqrt{|C_4(t+C_2)^2-1|}}+\dfrac l2)x+\dfrac1{t+C_2}y,$$

$$\mathbf{v}_3=(C_8(t+C_2)-\dfrac{C_4(t+C_2)}{2\sqrt{|C_4(t+C_2)^2-1|}}\arcsin\dfrac1{\sqrt{|C_4|}(t+C_2)}-\dfrac1{t+C_2})x+$$

$$+\dfrac{\omega_2(0)}{t+C_2}y+\omega_3(0).$$

To solve the problem of the gradient catastrophe we find the
initial values of velocity such as the function
$f(t)=(C_4(t+C_2)^2-1)(t+C_2)$ has positive zeroes. Zeroes of this
functions are $t=T_1=-C_2$ and
$t=T_{2,3}=\pm\dfrac1{\sqrt{C_4}}-C_2$. $T_2>0$ and $T_3>0$ if:

$$\begin{cases} \dfrac{\beta (0)}{\alpha (0)}+1>0,\\
\dfrac{(\beta (0))^{-3/2}}{\sqrt{(\alpha (0))^{-1}+(\beta
(0))^{-1}}}>\dfrac1{\beta (0)}.\end{cases}$$

Thus, we obtain:

1). If $\beta (0)>0$ and $\alpha (0)>-\beta (0)$ then gradient
catastrophe doesn't appear;

2). If $\beta (0)>0$ and $\alpha (0)<-\beta (0)$ then gradient
catastrophe appears in the time

$$t=\dfrac1{\beta (0)}\left(\sqrt{\dfrac{\alpha (0)}{\alpha
(0)+\beta (0)}}-1\right);$$

3). If $\beta (0)<0$ and $0<\alpha (0)<-\beta (0)$ then gradient
catastrophe appears in the time $T=-(\beta (0))^{-1}$;

4). If $\beta (0)<0$ and $\alpha (0)>-\beta (0)$ then gradient of
velocity turns in infinity in two points:

$$t=T_1=\dfrac1{\beta
(0)}\left(\sqrt{\dfrac{\alpha (0)}{\alpha (0)+\beta
(0)}}-1\right)$$

$$t=T_2=-(\beta (0))^{-1}$$

($T_1 < T_2$);

5). If $\beta (0)<0$ и $\alpha (0)<0$ then gradient of velocity
turns in infinity in three points:

$$t=T_1=\dfrac1{\beta (0)}\left(\sqrt{\dfrac{\alpha (0)}{\alpha
(0)+\beta (0)}}-1\right),$$

$$t=T_2=-(\beta (0))^{-1},$$

$$t=T_3=-\dfrac1{\beta (0)}\left(\sqrt{\dfrac{\alpha (0)}{\alpha
(0)+\beta (0)}}+1\right).$$

($T_1 < T_2 < T_3$).

\subsection{The case
$W(t,x,y)=\omega_1(t)x^2+\omega_2(t)xy+\omega_3(t)y^2+\lambda_1(t)x+$
$+\lambda_2(t)y+\lambda_3(t)$}

In this section the third component of velocity vector is function
$W(t,x,y)=$
$=\omega_1(t)x^2+\omega_2(t)xy+\omega_3(t)y^2+\lambda_1(t)x+\lambda_2(t)y+\lambda_3(t)$.
We get from (\ref{W(x,y)1}) and (\ref{W(x,y)2}):

\begin{equation}
\label{qv_W(x,y)}g(\dot{\omega}_1+2\alpha\omega_1+\xi \omega_2)=0,
\end{equation}

\begin{equation}
\label{qv_W(x,y)}g(\dot{\omega}_2+(\alpha +\beta)\omega_2+2\gamma
\omega_1+2\xi \omega_3)=0,
\end{equation}

\begin{equation}
\label{qv_W(x,y)}(\sigma+ g)(\dot{\omega}_3+2\beta\omega_3+\gamma
\omega_2)=0,
\end{equation}

\begin{equation}
\label{qv_W(x,y)}(2g+\sigma)(\dot{\lambda}_1+\alpha\lambda_1+\xi
\lambda_2)=0,
\end{equation}

\begin{equation}
\label{qv_W(x,y)}\sigma(\dot{\omega}_2+\alpha
+\beta\omega_2+2\gamma \omega_1+2\xi
\omega_3)+(g+\sigma)(\dot{\lambda}_2+\beta\lambda_2+\gamma
\lambda_1)=0,
\end{equation}

\begin{equation}
\label{qv_W(x,y)}g\dot{\lambda}_3+\sigma(\dot{\lambda}_1+\alpha\lambda_1+\xi
\lambda_2)=0.
\end{equation}

To solve this system consider several cases:

1. $g(t)\not\equiv 0$, $\sigma(t)\not\equiv 0$. Then we get from
(\ref{v0_1}) and (\ref{v0_2}) $\beta (t)\equiv 0$ and $\gamma
(t)\equiv 0$. Therefore, we solve the following system:

\begin{equation}
\label{qv_W(x,y)1_1}\dot{\omega}_1+2\alpha\omega_1+\xi\omega_2=0,
\end{equation}

\begin{equation}
\label{qv_W(x,y)2_1}\dot{\lambda}_1+\alpha\lambda_1+\xi
\lambda_2=0,
\end{equation}

\begin{equation}
\label{qv_W(x,y)3_1}\dot{\omega}_2+\alpha\omega_2+2\xi \omega_3=0,
\end{equation}

\begin{equation}
\label{qv_W(x,y)4_1}\dot{\omega}_3=0,
\end{equation}

\begin{equation}
\label{qv_W(x,y)5_1}\dot{\lambda_i}=0,\quad i=2,3.
\end{equation}

It easily follows from (\ref{qv_W(x,y)4_1})-(\ref{qv_W(x,y)5_1})
that $\omega_3(t)\equiv\omega_3(0)$,
$\lambda_i(t)\equiv\lambda_i(0)$, $i=2,3$.

Further, we find from (\ref{qv_W(x,y)2_1}) and
(\ref{qv_W(x,y)3_1}):

$$\omega_2(t)=\dfrac{2\lambda_2(0)}{\omega_3(0)}\lambda_1(t).$$

We have from (\ref{alpha})

\begin{equation}
\label{cond}\int \alpha (\tau)\,d\tau=\ln\sqrt{|2t+K_1|}.
\end{equation}

Using (\ref{xi0}) and (\ref{cond}), we can solve
(\ref{qv_W(x,y)1_1}) and (\ref{qv_W(x,y)2_1}):

$$\lambda_1(t)=-\lambda_2(0)\left(\dfrac l3(2t+K_1)+\dfrac{K_4t+K_7}{\sqrt{|2t+K_1|}}\right),$$

$$\omega_1(t)=\dfrac{\lambda_2(0)l}{18}(2t+K_1)^2+\dfrac{lK_7}{3}\sqrt{|2t+K_1|}+\dfrac{2\lambda_2(0)lK_4}{15}|2t+K_1|^{3/2}+$$

$$+\dfrac{\lambda_2(0)K_4t^2+2K_4K_7t}{2(2t+K_1)}+\dfrac{\lambda_2(0)lK_4}{3}t\sqrt{|2t+K_1|}+\dfrac{K_8}{2t+K_1},$$

\hspace*{-0,6cm} where

$$K_7=-\sqrt{|K_1|}\left(\dfrac{\lambda_1(0)}{\lambda_2(0)}+\dfrac{lK_1}3\right),$$

$$K_8=\omega_1(0)K_1-\dfrac1{18}\lambda_2(0)lK_1^3-\dfrac13lK_7|K_1|^{3/2}-\dfrac2{15}\lambda_2(0)lK_4|K_1|^{5/2}.$$

2. $g(t)\not\equiv 0$, $\sigma(t)\equiv 0$. We get from
(\ref{mass-conv3}) and (\ref{beta}) $\gamma (t)\equiv 0$ and $\beta
(t)=$ $\dfrac1{t+(\beta (0))^{-1}}$. Thus, we obtain the system:

\begin{equation}
\label{qv_W(x,y)1_2}\dot{\omega}_1+2\alpha\omega_1+\xi \omega_2=0,
\end{equation}

\begin{equation}
\label{qv_W(x,y)2_2}\dot{\lambda}_1+\alpha\lambda_1+\xi
\lambda_2=0,
\end{equation}

\begin{equation}
\label{qv_W(x,y)3_2}\dot{\omega}_2+(\alpha +\beta)\omega_2+2\xi
\omega_3=0,
\end{equation}

\begin{equation}
\label{qv_W(x,y)4_2}\dot{\lambda}_2+\beta\lambda_2=0,
\end{equation}

\begin{equation}
\label{qv_W(x,y)5_2}\dot{\omega}_3+2\beta\omega_3=0,
\end{equation}

\begin{equation}
\label{qv_W(x,y)6_2}\dot{\lambda}_3=0.
\end{equation}

Firstly, we get from (\ref{qv_W(x,y)6_2}) that
$\lambda_3(t)\equiv\lambda_3(0)$.

We have from (\ref{beta})

$$\int\limits_0^t \beta (\tau)\,d\tau=\ln|t+C_2|.$$

Therefore, it can be easily calculated that

\begin{equation}
\label{lambda_2}\lambda_2(t)=\lambda_2(0)\dfrac{1}{t+C_2},
\end{equation}

\begin{equation}
\label{omega_2}\omega_3(t)=\omega_3(0)\dfrac{1}{(t+C_2)^2}.
\end{equation}

Further, we have two cases:

a). If $\alpha (0)=0$ then from (\ref{a_&_x2}) $\alpha (t)\equiv
0$. For this reason we have the following system:

\begin{equation}
\label{qv_W(x,y)1_3}\dot{\lambda}_1+\xi \lambda_2=0,
\end{equation}

\begin{equation}
\label{qv_W(x,y)2_3}\dot{\omega}_1+\xi \omega_2=0,
\end{equation}

\begin{equation}
\label{qv_W(x,y)3_3}\dot{\omega}_2+\beta\omega_2+2\xi \omega_3=0,
\end{equation}

\hspace*{-0,6cm}where $\xi (t)=\dfrac{lt+C_3}{2(t+C_2)}$ (see
(\ref{xi2})).

Let us denote

\begin{equation}
\label{I}I(t)=\int\limits_0^t
\dfrac{l\tau+C_3}{2(\tau+C_2)^2}\,d\tau=l\ln|t+C_2|+\dfrac{lC_2-C_3}{t+C_2},
\end{equation}

\hspace*{-0,6cm}where $C_2=(\beta (0))^{-1},$ $C_3=2(\beta
(0))^{-1}\xi (0)$ (see (\ref{C})).

Using (\ref{I}), find the solution of (\ref{qv_W(x,y)1_3}) and
(\ref{qv_W(x,y)3_3}):



\begin{equation}
\label{lam1}\lambda_1(t)=-\dfrac{\lambda_2(0)}{2}(I(t)-C_9),
\end{equation}

\begin{equation}
\label{ome2}\omega_2(t)=-\omega_3(0)(I(t)-C_{10}),
\end{equation}

\hspace*{-0,6cm}where constants $C_9$ and $C_{10}$ are determined
from the initial data, namely:

$$C_9=\dfrac{2\lambda_1(0)}{\lambda_2(0)}+l\ln|C_2|+l-\dfrac{C_3}{C_2},$$

$$C_{10}=\dfrac{\omega_2(0)}{\omega_3(0)}+l\ln|C_2|+l-\dfrac{C_3}{C_2}.$$

Finally, we solve (\ref{qv_W(x,y)2_3}):


\begin{equation}
\label{ome1}\omega_1(t)=\dfrac{\omega_3(0)}{2}\Omega_1^0(t),
\end{equation}

\hspace*{-0,6cm}where

$$\Omega_1^0(u)=\dfrac12C_2l(1-l)\ln^2|u|+l^2u\ln|u|+C_{11}\ln|u|+(lC_{10}-l^2)u+\dfrac{C_{12}}{u}+C_{13},$$

\hspace*{-0,6cm}(here $u=t+C_2$).

$$C_{11}=C_2C_{10}(1-l)+l(lC_2-C_3),$$

$$C_{12}=C_2(l-1)(lC_2-C_3),$$

$$C_{13}=\dfrac{2\omega_1(0)}{\omega_3(0)}-\dfrac12C_2l(1-l)\ln^2|C_2|-(l^2C_2+C_{11})\ln|C_2|-(lC_{10}-l^2)C_2-\dfrac{C_{12}}{C_2}.$$

b). $\alpha (0)\neq0$, then $\alpha
(t)=((t+C_2)(C_4(t+C_2)^2-1))^{-1}$ (see (\ref{alpha2})) and
(\ref{C2}). We have:

$$-\int\limits_0^t \alpha (\tau)\,d\tau=\ln\dfrac{t+C_2}{\sqrt{|C_4(t+C_2)^2-1|}}.$$

$\lambda_1(t)$, $\omega_1(t)$ and $\omega_2(t)$ solve the system
(\ref{qv_W(x,y)1_2})-(\ref{qv_W(x,y)3_2}).

The equation (\ref{qv_W(x,y)1_2}) is the same as
(\ref{lin_W(x,y)1}). Therefore, we find:

$$\lambda_1(t)=L(t)\dfrac{t+C_2}{\sqrt{|C_4(t+C_2)^2-1|}}e^t,$$

\hspace*{-0,6cm}where

$$L(t)=C_{14}-\dfrac12C_4\arcsin\dfrac1{\sqrt{|C_4|}(t+C_2)}-\dfrac{\sqrt{|C_4(t+C_2)^2-1|}}{2(t+C_2)^2},$$

$$C_{14}=\dfrac{\sqrt{|C_4C_2^2-1|}}{C_2}(2C_2\lambda_1(0)-1)+\dfrac12C_4\arcsin\dfrac1{\sqrt{|C_4|}C_2}.$$

We find from (\ref{qv_W(x,y)3_2}):

$$\omega_2(t)=\Omega_2(t)\dfrac{1}{\sqrt{|C_4(t+C_2)^2-1|}},$$

\hspace*{-0,6cm}where the function $\Omega_2(t)$ can be found
from:

\begin{equation}
\label{Om}\Omega_2'(t)=-(l\sqrt{|C_4(t+C_2)^2-1|}+2C_5)\dfrac{\omega_3(0)}{(t+C_2)^2}.
\end{equation}

Let us denote $u=t+C_2$, $p(u)=\sqrt{C_4u^2-1}$. Further, we
integrate (\ref{Om}). Thus, we obtain:

$$\Omega_2(u)=\omega_3(0)\left(\dfrac{2C_5}{u}+\dfrac{l\sqrt{|C_4u^2-1|}}{u}-l\sqrt{|C_4|}\ln\mid\sqrt{C_4}u+\sqrt{|C_4u^2-1|}\mid+C_{15}\right),$$

\hspace*{-0,6cm}where

$$C_{15}=\dfrac{\omega_2(0)}{\omega_3(0)}\sqrt{|C_4C_2^2-1|}-\dfrac{2C_5+l\sqrt{|C_4C_2^2-1|}}{C_2}+
l\sqrt{|C_4|}\ln|2\sqrt{|C_4|}C_2|.$$

Finally, we have from (\ref{qv_W(x,y)2_2})

$$\omega_1(t)=W_1(t)\dfrac{(t+C_2)^2}{(C_4(t+C_2)^2-1)},$$

\hspace*{-0,5cm}where $W_1(t)$ can be found from

$$W_1'(t)=-(l\sqrt{|C_4(t+C_2)^2-1|}+2C_5)\dfrac{\Omega_2(t)}{(t+C_2)^2}.$$

It can be easy calculated that

$$W_1(u)=C_4\left(p(u)+u+\arctan{\dfrac1{p(u)}}+\ln|u|\right)+\sqrt{C_4}\ln|\sqrt{C_4}u+p(u)|+$$

$$+\dfrac{1-p(u)}{u}+\dfrac1{2u^2}+J(u),$$

\hspace*{-0,5cm}where

$$J(u)=\int\dfrac{p(u)(p(u)+1)}{u^2}\ln|\sqrt{C_4}u+p(u)|\,du.$$

Let us remark that conditions of the gradient catastrophe are the
same as for linear function $W(t,x,y)$.

\subsection{The case
$W(t,x,y)=\omega_1(t)x+\omega_2(t)y+\omega_3(t)z+\omega_4(t)$}

We will find the velocity vector and the density

\begin{equation}
\label{veloc}\textbf{v}=\begin{pmatrix} \alpha (t)x+\gamma(t)y\\
\xi(t)x+\beta (t)y\\W(t,x,y,z)
\end{pmatrix},\quad \rho(t,x)=\sigma(t)+g(t)x=\sigma(t)+\dfrac{\rho_0}{U}(\alpha
(t)+\beta (t))x.
\end{equation}

Here the divergency is

$$\alpha(t)+\beta(t)+\dfrac{\partial W}{\partial z},$$

the vorticity is

$$\left(\dfrac{\partial W}{\partial y}, -\dfrac{\partial
W}{\partial x}, \xi (t)+\gamma (t)\right)^T.$$

As early, we write the conservation of mass and the vorticity
conservation equation for our class of solution. Equation
(\ref{gas_syst1}) gives

$$\dot{\sigma}+\dot{g}x+(\alpha x+\gamma y)g+(\sigma+gx)(\alpha +\beta + \dfrac{\partial
W}{\partial z})=0.$$

Therefore

\begin{equation}
\label{m-conv1}\dot{g}+\alpha g+g(\alpha +\beta + \dfrac{\partial
W}{\partial z})=0,
\end{equation}

\begin{equation}
\label{m-conv2} \dot{\sigma}+\sigma(\alpha +\beta +
\dfrac{\partial W}{\partial z})=0,
\end{equation}

\begin{equation}
\label{m-conv3}g\gamma =0.
\end{equation}

From (\ref{vorticity}) we have the following equations:

$$\rho\left(\frac{\partial^2{W}}{\partial t\partial y}+\gamma
\frac{\partial{W}}{\partial x}+(\alpha x+\gamma
y)\frac{\partial^2{W}}{\partial x\partial y}+\beta
\frac{\partial{W}}{\partial y}+(\xi x+\beta
y)\frac{\partial^2{W}}{\partial
y^2}\right)+\frac{\partial{W}}{\partial
y}\frac{\partial{W}}{\partial z}+$$

\begin{equation}
\label{W(x,y,z)1}+W\frac{\partial^2{W}}{\partial y\partial
z}=\mu\left(\frac{\partial^3{W}}{\partial x^2\partial
y}+\frac{\partial^3{W}}{\partial
y^3}+\frac{\partial^3{W}}{\partial y\partial^2 z}\right),
\end{equation}

$$\rho\left(\frac{\partial^2{W}}{\partial t\partial x}+\alpha
\frac{\partial{W}}{\partial x}+(\alpha x+\gamma
y)\frac{\partial^2{W}}{\partial x^2}+\xi
\frac{\partial{W}}{\partial y}+(\xi x+\beta
y)\frac{\partial^2{W}}{\partial x\partial
y}+W\frac{\partial^2{W}}{\partial x\partial z}\right)+$$

$$+\frac{\partial{\rho}}{\partial
x}\left(\frac{\partial{W}}{\partial t}+(\alpha x+\gamma
y)\frac{\partial{W}}{\partial x}+(\xi x+\beta
y)\frac{\partial{W}}{\partial y}+W\frac{\partial{W}}{\partial
z}\right)=$$

\begin{equation}
\label{W(x,y,z)2}= -\mu\left(\frac{\partial^3{W}}{\partial
x^3}+\frac{\partial^3{W}}{\partial x\partial
y^2}\frac{\partial^3{W}}{\partial x\partial z^2}\right),
\end{equation}

$$\rho\left(\dot{\xi }-\dot{\gamma}+(\alpha +\beta )(\xi -\gamma )-l(\alpha +\beta )+m(\xi -\gamma )\right)+$$

\begin{equation}
\label{W(x,y,z)3}+\frac{\partial{\rho}}{\partial x}\left(\dot{\xi
}x+\dot{\beta }y+(\beta +m)(\xi x+\beta y)+(\xi -l)(\alpha
x+\gamma y)\right)=0.
\end{equation}

We consider the case
$W(t,x,y)=\omega_1(t)x+\omega_2(t)y+\omega_3(t)z+\omega_4(t)$. For
such function (\ref{W(x,y,z)1}) and (\ref{W(x,y,z)2}) imply:

\begin{equation}
\label{lin_W(x,y,z)1}(\sigma+gx)(\dot{\omega}_2+\gamma
\omega_1+\beta \omega_2+\omega_2\omega_3)=0,
\end{equation}

$$g(\dot{\omega}_1x+\dot{\omega}_2y+\dot{\omega}_3+(\alpha
x+\gamma y)\omega_1+(\xi x+\beta
y)\omega_2+\omega_3(\omega_1(t)x+\omega_2(t)y+$$

\begin{equation}
\label{lin_W(x,y,z)2}+\omega_3(t)z+\omega_4(t)))+(\sigma+gx)(\dot{\omega}_1+\alpha
\omega_1+\xi \omega_2+\omega_1\omega_3)=0
\end{equation}

(\ref{W(x,y,z)3}) is the same as (\ref{W(x,y)3}). Therefore, from
(\ref{W(x,y,z)3}) we obtain the system
(\ref{syst_v1})-(\ref{syst_v3}).

Further, we find the functions $\sigma(t)$, $\alpha(t)$,
$\beta(t)$, $\gamma (t)$, $\xi (t)$, $\omega_1(t)$, $\omega_2(t)$,
$\omega_3(t)$ and $\omega_4(t)$.

We can integrate the system
(\ref{lin_W(x,y,z)1})-(\ref{lin_W(x,y,z)2}) in the case
$g(t)\not\equiv 0$, $\sigma(t)\not\equiv 0$. We have from
(\ref{m-conv3}) $\gamma\equiv 0$. Then we may conclude from
(\ref{syst_v1}) and (\ref{syst_v3}) that $\beta\equiv 0$.

Thus, we have the system, which follows from
(\ref{lin_W(x,y,z)1})-(\ref{lin_W(x,y,z)2}) and facts
$\gamma\equiv 0$, $\beta\equiv 0$:

\begin{equation}
\label{vel_syst1}\dot{\omega}_1+(\alpha +\omega_3)\omega_1+\xi
\omega_2=0,
\end{equation}

\begin{equation}
\label{vel_syst2}\dot{\omega}_2+\omega_3\omega_2=0,
\end{equation}

\begin{equation}
\label{vel_syst3}\dot{\omega}_3+\omega_3^2=0,
\end{equation}

\begin{equation}
\label{vel_syst4}\dot{\omega}_4+\omega_3\omega_4=0,
\end{equation}

We solve the system (\ref{vel_syst2})-(\ref{vel_syst4}) and
obtain:

\begin{equation}
\label{sol3} \omega_3(t)=\dfrac1{t+c_3},
\end{equation}

\hspace*{-0,5cm}where $c_3=(\omega_3(0))^{-1}$, and

\begin{equation}
\label{sol2} \omega_i(t)=\dfrac{c_i}{t+c_3},
\end{equation}

\hspace*{-0,5cm}where $c_i=\omega_i(0)c_3$, $i=2,4$.
Here
$$g(t)=\dfrac{\rho_0}{U}(\alpha (t)+\beta (t))$$ (see
(\ref{veloc})). Therefore, we get from
(\ref{m-conv1})-(\ref{m-conv2}):

\begin{equation}
\label{m-c1}\dot{\alpha}+2\alpha^2+\omega_3\alpha=0.
\end{equation}

\begin{equation}
\label{m-c2} \dot{\sigma}+\sigma(\alpha+\omega_3)=0,
\end{equation}

Let us denote $A(t)=(\alpha(t))^{-1}$ and obtain from (\ref{m-c1})

$$\dot{A}-\omega_3A-2=0.$$

Now it can be calculated that:

\begin{equation}
\label{sol_alpha}\alpha(t)=(A(t))^{-1}=\dfrac{t+c_3}{t^2+2c_3t+c_5},
\end{equation}

\hspace*{-0,5cm}where $c_5=(\alpha(0))^{-1}c_3$.

We have from (\ref{sol_alpha}):

\begin{equation}
\label{int_alpha}\int\limits_0^t \alpha
(\tau)\,d\tau=\dfrac12\ln{|t^2+2c_3t+c_5|}.
\end{equation}

Using (\ref{int_alpha}), we solve (\ref{m-c2}):

\begin{equation}
\label{sol_sigma}\sigma(t)=\dfrac{c_6}{(t+c_3)\sqrt{t^2+2c_3t+c_5}},
\end{equation}

\hspace*{-0,5cm}where $c_6=\sigma(0)c_3\sqrt{c_5}$.

Further, we get from (\ref{syst_v3})

$$\dot{\xi}+\alpha\xi-l\alpha=0.$$

Consequently,

\begin{equation}
\label{sol_xi}\xi(t)=l+\dfrac{c_7}{\sqrt{t^2+2c_3t+c_5}},
\end{equation}

\hspace*{-0,5cm}where $c_7=(\xi(0)-l)\sqrt{c_5}$.

Finally, we find $\omega_1(t)$ from (\ref{vel_syst1})

\begin{equation}
\label{sol1}
\omega_1(t)=\dfrac{K(t)}{(t+c_3)\sqrt{t^2+2c_3t+c_5}},
\end{equation}

\hspace*{-0,5cm}where

$$K(t)=-c_2c_7t-\dfrac12c_2l\left((t+c_3)q(t)+(c_5-c_3^2)\ln|t+c_3+q(t)|\right)+c_1,$$

\hspace*{-0,5cm}(here $q(t)=\sqrt{t^2+2c_3t+c_5}$),

$$c_1=\omega_1(0)c_3\sqrt{c_5}+\dfrac12c_2l(c_3\sqrt{c_5}+(c_5-c_3^2)\ln|c_3+\sqrt{c_5}|).$$

Thus, in this case we obtain:

\begin{equation}
\label{r_z}\rho(t,x)=\dfrac{c_6}{(t+c_3)\sqrt{t^2+2c_3t+c_5}}+\dfrac{\rho_0(t+c_3)}{U(t^2+2c_3t+c_5)}x,
\end{equation}

\begin{equation}
\label{v(z)1}\mathbf{v}_1=\dfrac{t+c_3}{t^2+2c_3t+c_5}x,
\end{equation}

\begin{equation}
\label{v(z)2}\mathbf{v}_2=\left(l+\dfrac{c_7}{\sqrt{t^2+2c_3t+c_5}}\right)x,
\end{equation}

\begin{equation}
\label{v(z)3}\mathbf{v}_3=\dfrac{K(t)}{(t+c_3)\sqrt{t^2+2c_3t+c_5}}x+\dfrac{c_2}{t+c_3}y+\dfrac{1}{t+c_3}z+\dfrac{c_4}{t+c_3},
\end{equation}

\hspace*{-0,6cm}where the function $K(t)$ and the constants $c_i$,
$i=2,...,7$ are determined early.

To study whether the gradient catastrophe arises, we find the
positive zeroes of $f(t)=$ $=(t+c_3)(t^2+2c_3t+c_5).$ Let us denote
$B=c_3$, $-BC=c_5$. Thus we obtain the problem just considered in,
Section 1.

\end{document}